\documentclass[aps,prl,superscriptaddress,floatfix,twocolumn,showpacs]{revtex4-1}

\usepackage{verbatim}

\usepackage{amsmath}
\usepackage{amsfonts}
\usepackage{amssymb}
\usepackage{graphicx}
\usepackage{bm}
\usepackage{color}
\usepackage[hypertex]{hyperref}
\usepackage[active]{srcltx}
\usepackage{cases}
\usepackage{ulem}
\usepackage{multirow}

\newcommand{\be}{\begin{equation}}
\newcommand{\ee}{\end{equation}}
\newcommand{\bea}{\begin{eqnarray}}
\newcommand{\eea}{\end{eqnarray}}

\newcommand{\br}{{\bf r}}

\newcommand{\ex}{x}
\newcommand{\ci}{\mathfrak{i}}

\usepackage{wasysym}
\usepackage{graphicx}
\usepackage{bm}

\newcommand{\ColorOnline}{(Color online) }

\def\stacksymbols #1#2#3#4{\def\theguybelow{#2}
    \def\verticalposition{\lower#3pt}
    \def\spacingwithinsymbol{\baselineskip0pt\lineskip#4pt}
    \mathrel{\mathpalette\intermediary#1}}
\def\intermediary#1#2{\verticalposition\vbox{\spacingwithinsymbol
      \everycr={}\tabskip0pt
      \halign{$\mathsurround0pt#1\hfil##\hfil$\crcr#2\crcr
               \theguybelow\crcr}}}
\def\lapproxeq{\stacksymbols{<}{\sim}{2.5}{.2}}
\def\gapproxeq{\stacksymbols{>}{\sim}{3}{.5}}

\begin{document}
\title{Finite Size Effects and Irrelevant Corrections to Scaling near the Integer Quantum Hall Transition}

\author{Hideaki Obuse}
\altaffiliation{Present address: Department of Applied Physics, Hokkaido
University, Sapporo 060-8628, Japan}
\affiliation{Institute of Nanotechnology,
Karlsruhe Institute of Technology, 76344 Eggenstein-Leopoldshafen, Germany}
\author{Ilya A. Gruzberg}
\affiliation{The James Franck Institute and Department of Physics, The University of Chicago, Chicago, IL 60637, USA}
\author{Ferdinand Evers}
\affiliation{Institute of Nanotechnology, Karlsruhe Institute of Technology, 76344 Eggenstein-Leopoldshafen, Germany}
\affiliation{Institut f\"ur Theorie der kondensierten Materie, Universit\"at Karlsruhe, 76131 Karlsruhe, Germany}
\affiliation{Center for Functional Nanostructures, Karlsruhe Institute of Technology, 76131 Karlsruhe, Germany}

\date{May 9, 2012}

\begin{abstract}

We present a numerical finite size scaling study of the localization length in long cylinders near the integer quantum Hall transition (IQHT) employing the Chalker-Coddington network model. Corrections to scaling that decay slowly with increasing system size make this analysis a very challenging numerical problem. In this work we develop a novel method of stability analysis that allows for a better estimate of error bars. Applying the new method we find consistent results when keeping second (or higher) order terms of the leading irrelevant scaling field. The knowledge of the associated (negative) irrelevant exponent $y$ is crucial for a precise determination of other critical exponents, including multifractal spectra of wave functions. We estimate $|y| \gapproxeq 0.4$, which is considerably larger than most recently reported values. Within this approach we obtain the localization length exponent $2.62 \pm 0.06$ confirming recent results. Our stability analysis has broad applicability to other observables at IQHT, as well as other critical points where corrections to scaling are present.

\end{abstract}


\maketitle

Despite a long history, scaling properties of the integer quantum Hall transition (IQHT) still pose a considerable challenge \cite{rmp08}. In addition to the critical exponent $\nu$ of the localization length near the transition, recent studies \cite{Evers01,Obuse08,Evers08} analyze the multifractal spectrum $\Delta_q$ that describes the scaling of moments of the local density of states (LDOS) with the system size: $\langle \varrho({\br})^q \rangle \sim L^{-\Delta_q}$. This spectrum can, in principle, be measured in STM experiments. Though promising experiments have been undertaken \cite{morgenstern03, hashimoto08}, the presently attainable energy resolution appears to be insufficient to allow a sufficiently accurate measurement of $\Delta_q$. Hence, one relies on numerical simulations of critical wave functions statistics which relate to $\Delta_q$ via $ \langle (L^d |\psi({\br})|^2)^q \rangle \sim L^{-\Delta_q}$ \cite{rmp08}.

On the other hand, one can hope to calculate the spectrum $\Delta_q$ analytically. In the past decade, several proposals have been made for the quantum field theory underlying the IQHT critical point \cite{zirnbauer99, bhaseen00, tsvelik07, leclair07, luetken06, Bettelheim12}. However, most of them rely on assumptions that can neither be taken for granted nor easily checked against experiments. The identification of the correct critical theory remains an outstanding open problem, and predictions of the proposed models have to be taken with a grain of salt. For example, Wess-Zumino-type theories \cite{zirnbauer99, bhaseen00, tsvelik07, leclair07} are expected to exhibit a strictly parabolic multifractality spectrum $\Delta_q \propto q(1-q)$. However, the existing numerical evidence contradicts this prediction because $\Delta_q$ shows a significant, albeit small, quartic component $\sim(q-1/2)^4$ \cite{Obuse08, Evers08}.

Generally, numerical studies are performed for finite systems. How well the true asymptotic scaling regime can actually be probed, often depends crucially on a careful analysis of subleading corrections to scaling near a critical point. They tend to mask the true long-distance asymptotics and usually constitute the main difficulty for analysis of high-precision numerical simulations.

Two different kinds of subleading corrections should be distinguished. They can be easily understood in the framework of renormalization group (RG), and we adopt the corresponding terminology in the following discussion. One kind of subleading behavior is due to the fact that expectation values $\langle \ldots \rangle$ are calculated using a Hamiltonian (or an action functional) whose parameters (coupling constants) lie on a critical surface, but which does not coincide with the fixed point Hamiltonian. Deviations from the fixed point along the critical surface are parametrized by irrelevant scaling fields, whose decay under RG flow is described by irrelevant exponents $y_i < 0$. The leading irrelevant field decays with the exponent $y$ with the smallest absolute value. Its contributions are present in subleading corrections to scaling of any correlation function in a finite system at criticality.

The second kind of subleading corrections is associated with the specific correlation functions that one studies. The point is that a particular physical observable of interest may not exhibit pure scaling behavior even at the RG fixed point. Only certain carefully chosen observables correspond to {\it pure} scaling operators in the fixed point theory. As has recently been emphasized, moments of the LDOS and critical wave functions at an Anderson transition are examples of such pure scaling operators \cite{gruzberg11}. Thus, for moments of critical wave functions we expect corrections to scaling to come only from irrelevant fields, and primarily from the leading irrelevant field:
\be
\label{e1}
\big\langle (L^d|\psi({\br})|^2)^q \big\rangle = c_q L^{-\Delta_q}\big(1+ b_q L^{y} + b^\prime_q L^{2y} + \ldots\big).
\ee
Even though the correction terms ($L^{y},\ldots$) do not influence scaling to leading order (hence the term ``irrelevant''), they are still important to be studied because they ultimately determine the size of the critical parameter window (the true scaling regime) \cite{fn2}. As opposed to moments of LDOS, moments of the so-called point contact conductances are {\it not} pure scaling operators. We will return to this issue in subsequent papers.

For many Anderson transitions corrections to scaling decay sufficiently fast, so that the true asymptotic behavior is reliably addressed by a simple scaling analysis of numerical simulations \cite{rmp08, slevin99, kagalovsky99, asada02, rodriguez10, bondesan12}. The situation appears to be much less favorable for the IQHT. Over time it became clear that the Chalker-Coddington (CC) network model \cite{chalker88, kramerReview05}, specifically designed for numerical analysis, exhibits significant corrections to scaling which are very difficult to take into account systematically \cite{rmp08}.

The problem manifests itself in the recently reported low value $|y| \approx 0.17$ \cite{slevin09}, which is smaller than earlier estimates by more than a factor of two \cite{rmp08}. Recent  developments climaxed when Amado {\it et al.} \cite{Amado11} reported the presence of logarithmic corrections near the critical point in the CC model, implying, in a sense, $y=0$. Concomitant with this was an estimate for the ratio $\Gamma_c$ of the quasi-one dimensional localization length and the system width at the IQHT. Conformal invariance, which is expected to hold at the IQHT, predicts the relation $\Gamma_c/\pi = \alpha_0 - 2$, where $\alpha_0 = d\Delta_q/dq\big|_{q=0}$ \cite{Janssen94, Dohmen96, Obuse10}. The values $\alpha_0 - 2 = 0.2596$ and $0.2617$ obtained in Refs.\ [\onlinecite{Obuse08}, \onlinecite{Evers08}] are drastically different from the estimate $\Gamma_c/\pi=0.223 [0.219,0.228]$ reported in Ref. [\onlinecite{Amado11}]. Taken at face value, the discrepancy would indicate a breakdown of conformal invariance at the IQHT fixed point. In addition, if the value $|y|$ were indeed $0.17$ or smaller, the critical window for the IQHT would hardly be accessible with the currently attainable system sizes. As a consequence, reliable estimates for other critical exponents, like $\Delta_q$, could not be obtained.

\begin{figure}[tbp]
\includegraphics[width=0.85\columnwidth]{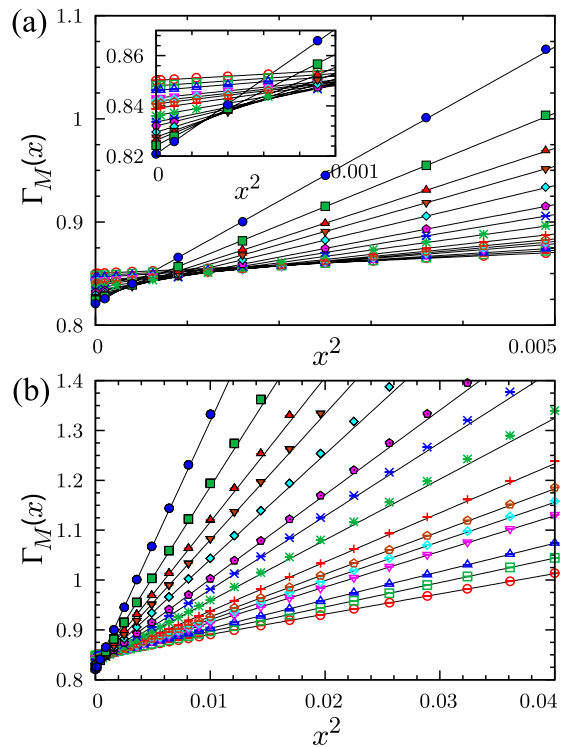}\\
\vspace{-0.3cm}
\caption{
\ColorOnline
(a) $\Gamma_M(x)$ as a function of $x^2$ for the circumference (number of links in the transverse direction) $M = 16, 20, 24, 32, 36, 40, 48, 64, 80, 96, 128, 160, 192, 256, 384$ (from the bottom to the top at $x^2 \approx 0.005$). The relative standard errors (one sigma) of our data are $0.005\%$ for $M\le64$, $0.02\%$ for $80\le M\le192$ and $0.05\%$ for $M=256, 384$. The solid lines are obtained from fitting the first two terms in Eq.\ (\ref{e2}) to the data for each $M$. Inset: Enlargement of $\Gamma_M(x)$ near $x=0$. (b) Same as (a) but
in a wider range of $x^2$.
}
\label{f1}
\vspace{-0.5cm}
\end{figure}

Motivated by this situation we revisit the finite size scaling corrections to the localization length in quasi-one dimensional geometry (Q1D) near the IQHT. In addition to including the leading irrelevant scaling field in our analysis, we develop and employ a new method of ``stability map'' in the parameter space. The fitting procedure adjusts two critical exponents (the leading irrelevant exponent $y$ and the exponent of the localization length $\nu$) and four or more coefficients in the expansion of the scaling function. Despite many fitting parameters, reliability of our results is demonstrated by the stability map analysis. As a result, we arrive at the following conclusions: the estimate of the critical exponent for the localization length $\nu= 2.62\pm0.06$ obtained by previous authors \cite{slevin09, Obuse10, Amado11, Dahlhaus11, fulga11, slevin12} is confirmed. Furthermore, we find $|y| \gapproxeq 0.4$ and $\Gamma_c/\pi = 0.257\pm0.002$. The latter value is consistent with predictions coming from conformal invariance, as well as earlier estimates \cite{Obuse08, Evers08, Evers01}.

To calculate the localization length $\xi_M(\ex)$ in Q1D we employ the isotropic version of the CC network model \cite{chalker88, kramerReview05} on cylinders (meaning, with periodic boundary conditions in one direction) of length $L$ and circumference $M$. $\xi_M(x)$ is calculated from the standard transfer matrix method \cite{mackinnon83, slevin09}. The relevant scaling field $x$ parametrizes the non-random part of the transfer matrix at a node as
\begin{math}
\begin{pmatrix}
t^{-1} & rt^{-1} \\ rt^{-1} & t^{-1}
\end{pmatrix}
\end{math},
where $t^{-2} = e^{2x} + 1$ and $r^2=1-t^2$.

To extract critical exponents we fit numerical data for the finite size ratio $\Gamma_M(x) \equiv M/\xi_M(x)$ by the following expression for the scaling function:
\begin{align}
\label{e2}
\Gamma(\ex,M) &= \gamma(M) + \ex^2 \gamma^\prime(M) + {\mathcal O}(\ex^4M^{4/\nu}),\\
\label{e3}
\gamma(M) &= \Gamma_c \big( 1 + a_1 M^{y} + a_2 M^{2y} + \ldots \big),  \\
\label{e4}
\gamma^\prime(M) &= \Gamma^\prime M^{2/\nu} \big(1 + a^\prime_1 M^{y} {+} a_2^\prime M^{2y}  {+} \ldots\big).
\end{align}
(Notice that we distinguish the data $\Gamma_M(x)$ from the scaling function $\Gamma(x,M)$ by using different notation.) The analytical form of Eqs. (\ref{e2}--\ref{e4}) is standard for a scaling function near a critical fixed point combined with a fact that $\Gamma(x,M)$ is an even function of $x$ at the IQHT \cite{slevin09}. It involves an expansion in even powers of $x$ and integer powers of the leading irrelevant field that itself is assumed to scale as $M^{y}$ with the systems width. We keep only the leading order in $\ex^2M^{2/\nu}$ to reduce the number of fitting parameters. Moreover, while we will keep the terms up to $M^{3y}$ for $\gamma(M)$, we will ignore terms starting from $M^{2y}$ for $\gamma^\prime(M)$. The reason is that within our range of system sizes keeping these correction terms changes the exponent $\nu$ at the level of less than one percent.

\begin{figure}[tbp]
\includegraphics[width=0.90\columnwidth]{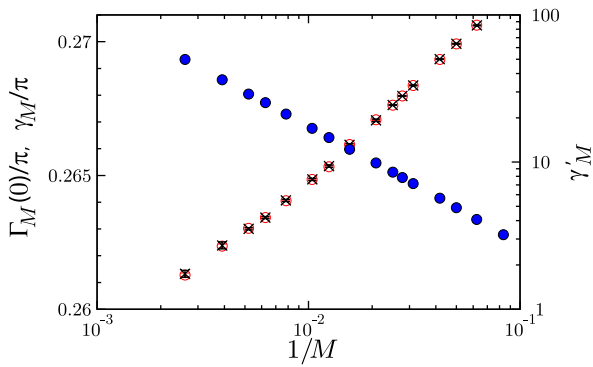}
\vspace{-0.3cm}
\caption{
\ColorOnline
Original data $\Gamma_M(x=0)$ (black $\times$, left axis) and the offset $\gamma_M$ (red $\circ$, left axis) and the slope $\gamma^\prime_M$ (blue $\bullet$, right axis) obtained from fitting the first two terms in Eq. (\ref{e2}) to $\Gamma_M(x)$ for $x \le 0.05$.
}
\label{f2}
\vspace{-0.5cm}
\end{figure}

We show numerical data $\Gamma_M(x)$ in Fig. \ref{f1}. The solid straight lines in the figure show the results of the least square fit of the first two terms in Eq. (\ref{e2}) to the data, performed separately for each $M$ in the range $x \le 0.05$. We will denote the offset and the slope of the straight lines obtained in this way by $\gamma_M$ and $\gamma'_M$. We see from the plots that corrections coming from $x^4$ (and higher order terms) are very small even within the range of $x^2$ shown in Fig.\ \ref{f1}(b). This justifies keeping only terms up to order $x^2$ in the expansion (\ref{e2}). The effects of corrections to scaling are most pronounced near the critical point (Fig. \ref{f1}, inset) where the solid lines intersect at nonzero values of $x$ instead of meeting at a single point $\Gamma_c$ at $x=0$.

\begin{figure*}[t]
\includegraphics[width=\textwidth]{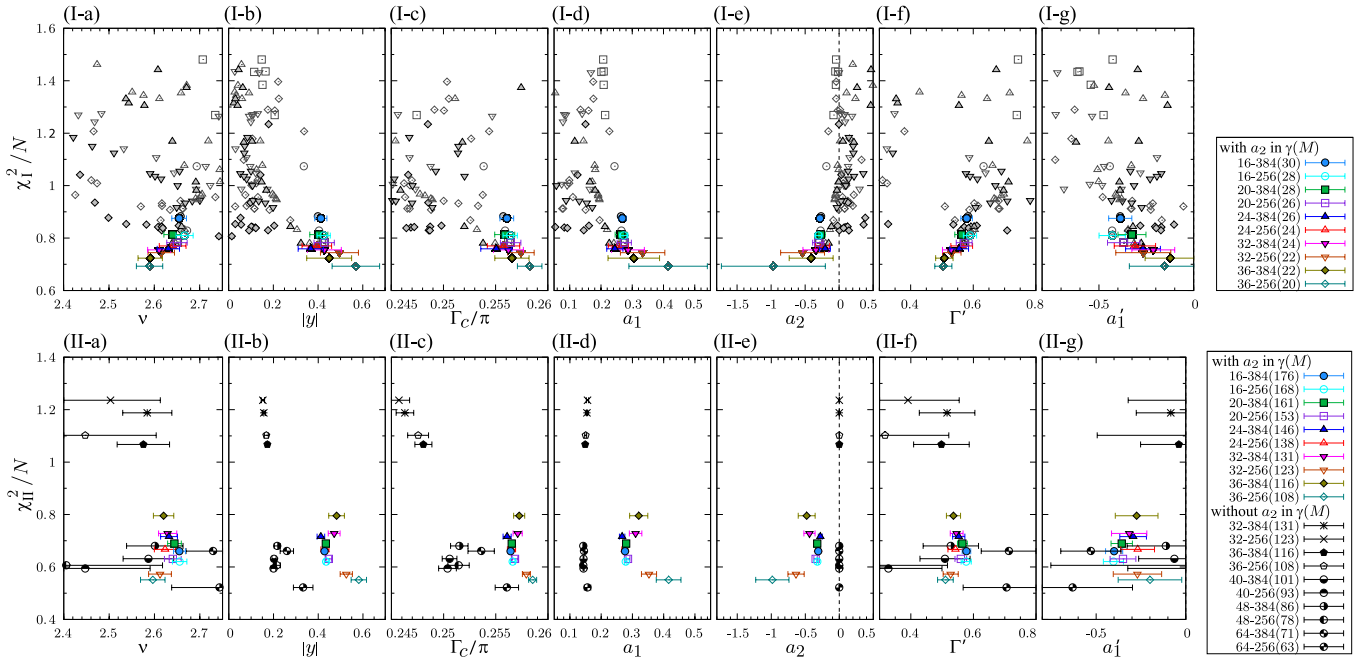}
\caption{
\ColorOnline
Stability map for the exponents $\nu$, $|y|$ and the coefficients $\Gamma_c/\pi$,  $a_1$, $a_2$, $\Gamma'$, $a^\prime_1$ appearing in Eqs. (\ref{e3}, \ref{e4}), obtained by methods I and II for $0 \le x\le 0.05$. In a given panel each symbol shows a local minimum of the cost function $\chi^2_\text{I,II}$ divided by the number $N$ of data points taken into account. Different symbols represent different input data sets obtained by varying the range of system sizes $M$ and the number of data points $N$, and listed in the legend boxes at the right as $M_\text{min}$--$M_\text{max} (N)$. Upper panel: In the case of $\chi^2_\text{I}$ many local minima (shown by gray symbols) have been found. The global minima of $\chi^2_\text{I}$ for each input data set are shown by colored symbols with error bars \cite{error}. Lower panel: For $\chi^2_\text{II}$ a single minimum (colored symbols) was found for each data set, when the term $a_2 M^{2y}$ in Eq. (\ref{e3}) was included in the analysis. We also show results of minimization (black symbols) for the case when a single correction term $a_1 M^{y}$ in $\gamma(M)$ was taken into account for input data with $M_\text{min}\le 64$. On the panel (II-e) the black symbols are shown on the vertical line $a_2 = 0$. We note that since $\chi^{2}/N$ tends to be small for small $N$, a subtle difference of $\chi^2$ between different input data sets is meaningless.
} \label{f3}
\vspace{-0.5cm}
\end{figure*}

In Fig.~\ref{f2} we plot the offset $\gamma_M$ and the slope $\gamma_M^\prime$ obtained from fitting by Eq. (\ref{e2}), together with the original data for $\Gamma_M(x=0)$. We see that the all these quantities significantly depend on the system size $M$. The variation of $\Gamma_M(0)$ over the available range of $M$ is $\sim4\%$. The very slow (approximately logarithmic) dependence on $M$ seen in Fig. \ref{f2} constitutes the notorious difficulty for numerical studies on the IQHT mentioned in the introduction. At this point we can proceed in two different ways.

{\it Method I (the two-step optimization method).} In this method we take a subset of $\Gamma_M(0)$ and $\gamma'_M$ presented in Fig. \ref{f2} and use these as the input data set for fitting the functions $\gamma(M)$ and $\gamma'(M)$ from Eqs. (\ref{e3}, \ref{e4}). These functions depend on a number of fitting parameters $\nu$, $y$, $\Gamma_c$, $\Gamma'$, $a_1$, $a^\prime_1$, etc., that we optimize by finding a minimum to the ``cost'' function
\be
\chi^2_\text{I} = \sum_{M} \frac{(\Gamma_M(0) -\gamma(M))^2}{\sigma_{M}^2(0)} + \sum_{M} \frac{(\gamma_{M}^\prime{-}\gamma^\prime(M))^2}{{\sigma^\prime_{M}}^2},
\ee
where $\sigma_{M}(0),\sigma^\prime_{M}$ are the (absolute) statistical errors of the input data $\Gamma_M(0)$, $\gamma^\prime_{M}$ \cite{method-I}.

{\it Method II (the global optimization method).} In this method we take a subset of all numerical data $\Gamma_M(x)$ as the input for a global fitting by the function $\Gamma(x,M)$ from Eqs. (\ref{e2}-\ref{e4}). Optimal parameters of $\Gamma(x,M)$ are found by minimizing the cost function
\be
\chi^2_\text{II} = \sum_{M} \sum_{x} \frac{(\Gamma_M(x)-\Gamma(x,M))^2}{\sigma^2_{M}(x)},
\ee
where $\sigma_M(x)$ are the (absolute) standard errors of
$\Gamma_M(x)$.

In both methods we can vary the input sets by varying the range $M_\text{min}$--$M_\text{max}$ of the system sizes and the total number $N$ of data points in the set.

We have employed both methods I and II with more than 1000 fitting trials for each input data set, starting with different initial fitting parameters chosen randomly \cite{fitting}. To visualize the results, we show a ``stability map'' in Fig.\ \ref{f3}. It displays minima of $\chi^2_\text{I,II}$ as functions of all fitting parameters. The map allows for a better estimate of error bars and helps understanding how fitting parameters affect the cost functions.

For method II the optimization routine yields a unique minimum of $\chi^2_\text{II}$ for each data set. The minima for sets with different ranges of $M$ and different numbers of points $N$ are tightly clustered together if we take into account the terms up to $a_2 M^{2y}$ in the function $\gamma(M)$, but not when we keep only the first correction term $a_1 M^{y}$, see lower panels in Fig. \ref{f3}.

In stark contrast, the cost function $\chi^2_\text{I}$ exhibits multiple local minima where method I gets stuck. On the upper panels in Fig. \ref{f3} we show the local minima with low values of $\chi^2_\text{I}$ obtained by method I with terms up to $a_2 M^{2y}$ in $\gamma(M)$. From the panel (I-b) one infers that the local minima exist in a wide range $0 < |y| < 0.7$.
However, only for values of $|y|$ above $|y| \approx 0.4$ the cost function $\chi^2_\text{I}$ approaches its global minimum, which is emphasized by colored symbols with error bars. At this global minimum the results of methods I and II agree.

From the fitted data presented in Fig. \ref{f3} we extract several observations \cite{suppl}:

(i) If only the $a_1 M^{y}$ term is included in $\gamma(M)$, and $M_\text{min}\le 40$, we find $|y| \lapproxeq 0.2$. This is consistent with a previously reported value \cite{slevin09}. The small apparent exponent $y$ reflects the slow decay of $\gamma_M$ with $1/M$ seen in Fig. \ref{f2}. Remarkably, when the next correction term is kept in $\gamma(M)$,  the optimal value for $|y|$ increases; we obtain $0.4 \apprle |y| \apprle 0.6$ \cite{a2=0}. This estimate remains unchanged when we add the term $a_3 M^{3y} $\cite{suppl}. Larger values of $|y|$ obtained when $a_{1,2}$-terms and higher ones are included do not contradict the slow decay of $\gamma_{M}$ if these terms partially cancel each other for small system sizes. Indeed, we see from Fig. \ref{f3} that the coefficients $a_1$ and $a_2$ are similar in magnitude but opposite in sign.

(ii) The estimate of
$\Gamma_c/\pi = \alpha_0 - 2$ is sensitive to the number of correction terms kept
in $\gamma(M)$. If only the $a_1$ term is kept, and $M_\text{min}\le 40$, our value of $\Gamma_c$ is again consistent with the result $\alpha_0 - 2 \approx 0.248 [0.244,0.251]$ found earlier \cite{slevin09}. However, adding higher order terms gives a significantly larger value $\alpha_0 - 2 \approx 0.257 \pm 0.002$. This latter estimate is broadly compatible with results based on the wave function statistics \cite{Obuse08, Evers08, Evers01}.

(iii) We confirm the estimate for the localization length exponent $2.62 \pm 0.06$ \cite{nu} already obtained previously by several authors \cite{slevin09, Obuse10, Amado11, Dahlhaus11, fulga11}.

Given that finite size effects are quite significant near the IQHT, and that they have been treated very differently by previous authors, it is, perhaps, remarkable that similar estimates for $\nu$ were recently found. Using the same model as in this work, Ref. [\onlinecite{slevin09}] kept only a single $M^{y}$ term, and Ref. [\onlinecite{Amado11}] employed powers of $1/\ln M$. Different boundary conditions or models were employed in Refs.\ [\onlinecite{Obuse10}, \onlinecite{Dahlhaus11}], where only a single $M^{y}$ term was used. The authors of Ref.\ [\onlinecite{fulga11}] found corrections to scaling to be insignificant. We believe that this relative insensitivity of the estimates for $\nu$ to subleading corrections is related to the facts that (a) $\nu$ is large compared to $|y|$ and (b) the ranges of system sizes $M$ used are narrow, with the ratio $M_\text{max}/M_\text{min}$ hardly exceeding ten.

By the same reasoning, it is not clear why earlier estimates of $\nu$ obtained for the random-Landau-matrix model ($\nu=2.35\pm 0.03$ \cite{huckestein95}) or other
models ($\nu=2.33\pm0.09$ \cite{sandler04, slevin11}), differ from more recent values by up to 10\%. While a violation of universality at the IQHT would be a logically possible explanation, the present numerical evidence is not strong enough to draw such a drastic conclusion. We expect that the apparent discrepancy in estimates of scaling exponents obtained from different
models of the IQHT will disappear upon reinvestigating finite size effects more carefully.

{\it Conclusions.} Using the CC network model on long cylinders, we have numerically analyzed corrections to scaling near the IQHT. Our data is consistently interpreted using the standard form of corrections to scaling if more than one correction term is included in the scaling functions. The interpretation is facilitated by the ``stability map'' analysis. Our results satisfy predictions of conformal invariance.

We thank S.\ Bera, A.\ Furusaki, A.~D.\ Mirlin, T.\ Ohtsuki, A.\
Sedrakyan, K.\ Slevin and A.\ Tsvelik for numerous discussions of
corrections to scaling. We thank K.\ Slevin and T.\ Ohtsuki for sharing
Ref.\ \cite{slevin12} prior to publication, and A.\ W.\ W.\ Ludwig for
helpful comments. We also thank I. Kondov (SCC) for support and the
Juelich Supercomputer Center (project HKA12) for allocation of computing
time on JUROPA. Numerical simulations were performed in part using the
PADS resource (NSF Grant No.\ OCI-0821678) at the Computational Institute at the University of Chicago. H.\ O.\ is supported by a Grant-in-Aid for Research Abroad from JSPS. I.\ A.\ G.\ was supported by NSF Grants No. DMR-1105509 and No. DMR-0820054.

\clearpage
\setcounter{equation}{0}
\setcounter{table}{0}
\setcounter{figure}{0}
\setcounter{page}{1}

{\large \bf{Supplemental Material for ``Finite Size Effects and Irrelevant Corrections to Scaling near the Integer Quantum Hall Transition''}}

{\it
In this supplemental material, we present all fitting parameters for scaling functions with up to three correction terms, for various input data sets. We also show the stability maps for these parameters, and provide a table summarizing and comparing our results for critical exponents with those of other publications.
}

\section{scaling analysis}

We calculate the localization length $\xi_M(x)$ for the Chalker-Coddington network model in the quasi-one dimensional cylinder geometry close to the integer quantum Hall transition (at $x=0$). We use the following values for the circumference of the cylinder $M = 16$, $20$, $24$, $32$, $36$, $40$, $48$, $64$, $80$, $96$, $128$, $160$, $192$, $256$, $384$. To investigate the critical behavior, we employ the following scaling function for $\Gamma_M(x) \equiv M/\xi_M(x)$,
\begin{align}
\label{eq:scaling1}
\Gamma(\ex,M) &= \gamma(M) + \ex^2 \gamma^\prime(M),\\
\label{eq:scaling2}
\gamma(M) &= \Gamma_c\left( 1 + a_1 M^{y} + a_2 M^{2y} + a_3 M^{3y} \right),  \\
\label{eq:scaling3}
\gamma^\prime(M)&= \Gamma^\prime M^{2/\nu} \left(1 + a^\prime_1 M^{y} \right),
\end{align}
where the fitting parameters $\nu$ and $y$ represent the critical exponent of the localization length and the leading irrelevant exponent, respectively. The other fitting parameters are the coefficients $\Gamma_c, a_1, a_2, a_3, \Gamma^\prime, a_1^\prime$ of the scaling functions.

As is explained in the main paper, we consider two fitting methods. In method I we use the data set of $\Gamma_M(x=0)$, and $\gamma_M^\prime$ obtained from fitting Eq. (\ref{eq:scaling1}) to the data for $\Gamma_M(x)$. In method II we use the original data $\Gamma_M(x)$. Furthermore, in order to check that our results remain unchanged upon variations in  the $x$ range used in the fitting, we consider two ranges: $0\le x \le x_u$ with $x_u=0.05$ and $x_u=0.07$. Thereby, we have four kinds of input data sets total.

For each method, we seek the global minimum of the corresponding cost function $\chi^2$ (see Eqs.\ (5--6) in the main paper). We examine the stability of fitting with respect to keeping different number of correction terms in in Eq.\ (\ref{eq:scaling2}), from one to three. We also vary the minimum and maximum circumference, $M_\text{min}$ and $M_\text{max}$, of the input data sets to check stability of our fits.

\begin{figure*}[h]
\includegraphics[width=\textwidth]{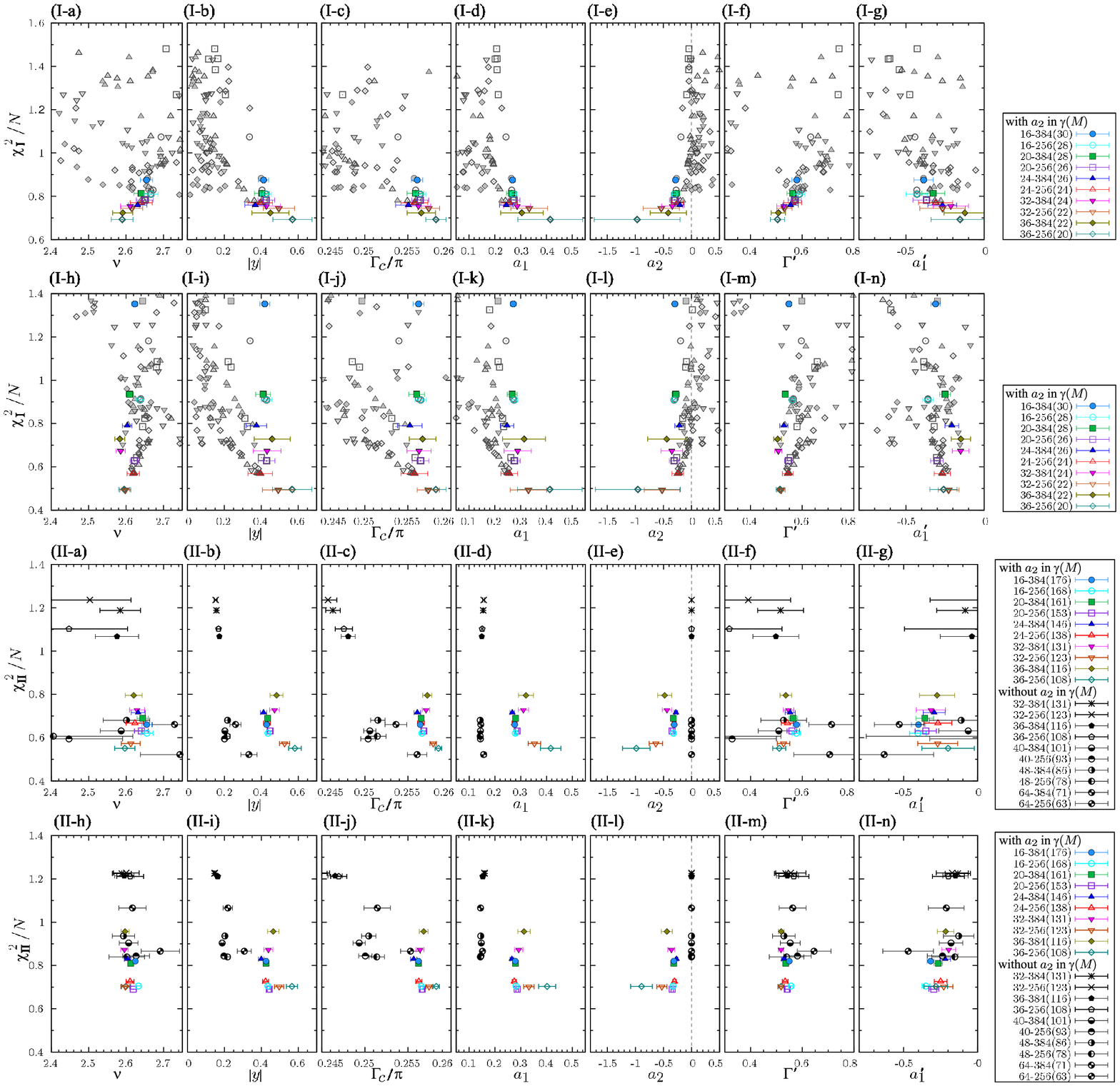}\\
\caption{
Stability maps for the exponents $\nu, |y|$ and the coefficients $\Gamma_c/\pi,  a_1,a_2,\Gamma',a^\prime_1$. From top to bottom: method I with $x_u=0.05$ [(I-a)-(I-g)], method I with $x_u=0.07$ [(I-h)-(I-n)], method II with $x_u=0.05$ [(II-a)-(II-g)], and method II with $x_u=0.07$ [(II-h)-(II-n)].
}
\label{fig:stability}
\end{figure*}

Tables \ref{t1} - \ref{t4} shows results of fitting corresponding to the global minimum of $\chi^2$ for the four kinds of input data, {\it i.e.}, method I with $x_u= 0.05$, method I with $x_u= 0.07$, method II with $x_u= 0.05$, and method II with $x_u= 0.07$. In addition, we show the stability maps for these input data in Fig.\ \ref{fig:stability}.

Some remarks are in order.

(i) Even if the input data for $x_u= 0.07$ are used, fitting parameters for the global minimum are consistent with those obtained from the input data for $x_u= 0.05$. Moreover, distributions of the local minima obtained by method I as shown in Fig.\ \ref{fig:stability} are also similar. For example, the local minima for $a_2$ locate near $a_2=0$, and most of those for $|y|$ are located below $|y| = 0.2$, etc. This demonstrates stability of fitting procedure with respect to changes in $x_u$.

(ii) In addition to the first two correction terms, we have examined the effect of the third correction term $a_3 M^{3y}$ in the scaling function $\gamma(M)$. Since the higher order terms can contribute appreciably to $\Gamma_M(x)$ only for small $M$, we fix $M_\text{min}$  to a small value, $16$. (For larger values of $M_\text{min}$ results of the fitting become unreliable, which manifests itself in error bars of fitting parameters exceeding their mean values.) Keeping the term $a_3 M^{3y}$ in the scaling function results in changes of all fitting parameters by only a few percent. In contrast, when we go from one correction term in the scaling function to two (up to $a_2 M^{2y}$), some fitting parameters change drastically (for example, $|y|$ increases by a factor of 3, the coefficient $a_1$ grows by a factor of 2, etc.). We also observe that the signs of $a_1$ and $a_2$ are opposite when including the second correction term, and this feature remains unchanged when adding the third correction term to the analysis. As we have explained in the main paper, the opposite signs lead to the estimate of $|y|$ to grow due to partial cancellations. These results imply that keeping two correction terms in $\gamma(M)$ is sufficient for our purposes.

\section{comparison with previous works}

For readers' convenience, we compare our results for $\nu$, $y$, and $\Gamma_c/\pi$
($=\alpha_0-2$) with those of other authors in Table \ref{tab:compare}.

\clearpage


\begin{table*}
\rotatebox{90}{
\begin{minipage}{1\textheight}
\caption{
List of fitting parameters corresponding to the global minimum $\chi^2_\text{min}$ of the cost function, obtained by method I for $0\le x \le0.05$. $M_\text{min}$ and  $M_\text{max}$ represent the minimum and maximum circumferences, respectively, used in the analysis. $N$ and $Q$ indicate the total number of input data points and goodness of fit, respectively.
\label{t1}
}
\begin{tabular}{c c c|c c | c c c | c c c| c c}
\hline\hline
$M_\text{min}$ &  $M_\text{max}$ & $N$ & $\chi^2_\text{min}/N$ & $Q$
 &$\nu$ & $|y|$ & $\Gamma_c/\pi \equiv \alpha_0-d$ &
$a_1$ &$a_2$  &$a_3$ & $\Gamma^\prime$ & $a_1^\prime$\\
\hline
$24$ & $384$ & $26$ & $1.008$ & $0.159$ & $2.712 \pm 0.055$ & $0.136 \pm 0.012$  & $0.24349 \pm 0.00203$ & $0.1639 \pm 0.0078$ & $-$ & $-$ & $0.768 \pm 0.139$ & $-0.442 \pm 0.203$ \\
$32$ & $384$ & $24$ & $0.935$ & $0.213$ & $2.669 \pm 0.059$ & $0.162 \pm 0.020$  & $0.24699 \pm 0.00231$ & $0.1519 \pm 0.0073$ & $-$ & $-$ & $0.657 \pm 0.120$ & $-0.329 \pm 0.214$ \\
$36$ & $384$ & $22$ & $0.828$ & $0.311$ & $2.622 \pm 0.062$ & $0.181 \pm 0.024$  & $0.24886 \pm 0.00225$ & $0.1470 \pm 0.0058$ & $-$ & $-$ & $0.568 \pm 0.104$ & $-0.190 \pm 0.227$ \\
$40$ & $384$ & $20$ & $0.680$ & $0.480$ & $2.567 \pm 0.067$ & $0.200 \pm 0.028$  & $0.25035 \pm 0.00210$ & $0.1442 \pm 0.0039$ & $-$ & $-$ & $0.481 \pm 0.094$ & $0.009 \pm 0.255$ \\
$48$ & $384$ & $18$ & $0.645$ & $0.477$ & $2.548 \pm 0.068$ & $0.243 \pm 0.043$  & $0.25281 \pm 0.00210$ & $0.1445 \pm 0.0025$ & $-$ & $-$ & $0.457 \pm 0.085$ & $0.088 \pm 0.270$ \\
$64$ & $384$ & $16$ & $0.497$ & $0.634$ & $2.651 \pm 0.086$ & $0.255 \pm 0.064$  & $0.25333 \pm 0.00274$ & $0.1457 \pm 0.0067$ & $-$ & $-$ & $0.599 \pm 0.133$ & $-0.308 \pm 0.325$ \\
\hline
$24$ & $256$ & $24$ & $1.017$ & $0.142$ & $2.749 \pm 0.070$ & $0.132 \pm 0.013$  & $0.24270 \pm 0.00234$ & $0.1670 \pm 0.0093$ & $-$ & $-$ & $0.851 \pm 0.195$ & $-0.512 \pm 0.254$ \\
$32$ & $256$ & $22$ & $0.993$ & $0.148$ & $2.701 \pm 0.080$ & $0.158 \pm 0.023$  & $0.24649 \pm 0.00277$ & $0.1536 \pm 0.0091$ & $-$ & $-$ & $0.717 \pm 0.175$ & $-0.403 \pm 0.282$ \\
$36$ & $256$ & $20$ & $0.908$ & $0.199$ & $2.635 \pm 0.087$ & $0.181 \pm 0.029$  & $0.24883 \pm 0.00265$ & $0.1471 \pm 0.0069$ & $-$ & $-$ & $0.589 \pm 0.148$ & $-0.228 \pm 0.308$ \\
$40$ & $256$ & $18$ & $0.745$ & $0.340$ & $2.544 \pm 0.098$ & $0.205 \pm 0.034$  & $0.25076 \pm 0.00240$ & $0.1434 \pm 0.0041$ & $-$ & $-$ & $0.450 \pm 0.129$ & $0.098 \pm 0.372$ \\
$48$ & $256$ & $16$ & $0.653$ & $0.402$ & $2.500 \pm 0.102$ & $0.273 \pm 0.054$  & $0.25415 \pm 0.00212$ & $0.1463 \pm 0.0062$ & $-$ & $-$ & $0.404 \pm 0.110$ & $0.302 \pm 0.415$ \\
$64$ & $256$ & $14$ & $0.427$ & $0.650$ & $2.718 \pm 0.142$ & $0.333 \pm 0.093$  & $0.25596 \pm 0.00235$ & $0.1588 \pm 0.0237$ & $-$ & $-$ & $0.671 \pm 0.206$ & $-0.557 \pm 0.531$ \\
\hline
$24$ & $192$ & $22$ & $1.020$ & $0.130$ & $2.755 \pm 0.076$ & $0.127 \pm 0.013$  & $0.24185 \pm 0.00258$ & $0.1705 \pm 0.0105$ & $-$ & $-$ & $0.873 \pm 0.221$ & $-0.525 \pm 0.278$ \\
$32$ & $192$ & $20$ & $1.027$ & $0.114$ & $2.705 \pm 0.090$ & $0.150 \pm 0.024$  & $0.24546 \pm 0.00322$ & $0.1572 \pm 0.0112$ & $-$ & $-$ & $0.734 \pm 0.205$ & $-0.416 \pm 0.320$ \\
$36$ & $192$ & $18$ & $0.957$ & $0.141$ & $2.630 \pm 0.099$ & $0.171 \pm 0.030$  & $0.24791 \pm 0.00312$ & $0.1497 \pm 0.0090$ & $-$ & $-$ & $0.584 \pm 0.173$ & $-0.216 \pm 0.356$ \\
$40$ & $192$ & $16$ & $0.785$ & $0.249$ & $2.516 \pm 0.117$ & $0.196 \pm 0.036$  & $0.25006 \pm 0.00281$ & $0.1448 \pm 0.0057$ & $-$ & $-$ & $0.413 \pm 0.147$ & $0.217 \pm 0.453$ \\
$48$ & $192$ & $14$ & $0.697$ & $0.283$ & $2.451 \pm 0.126$ & $0.264 \pm 0.059$  & $0.25378 \pm 0.00247$ & $0.1456 \pm 0.0053$ & $-$ & $-$ & $0.350 \pm 0.127$ & $0.548 \pm 0.524$ \\
$64$ & $192$ & $12$ & $0.492$ & $0.434$ & $2.739 \pm 0.184$ & $0.325 \pm 0.104$  & $0.25574 \pm 0.00279$ & $0.1568 \pm 0.0247$ & $-$ & $-$ & $0.702 \pm 0.277$ & $-0.609 \pm 0.662$ \\
\hline\hline
$16$ & $384$ & $30$ & $0.875$ & $0.289$ & $2.655 \pm 0.016$ & $0.413 \pm 0.027$  & $0.25607 \pm 0.00066$ & $0.2687 \pm 0.0116$ & $-0.2843 \pm 0.0417$ & $-$ & $0.579 \pm 0.018$ & $-0.387 \pm 0.061$ \\
$20$ & $384$ & $28$ & $0.813$ & $0.356$ & $2.640 \pm 0.019$ & $0.404 \pm 0.040$  & $0.25587 \pm 0.00096$ & $0.2642 \pm 0.0189$ & $-0.2688 \pm 0.0648$ & $-$ & $0.563 \pm 0.022$ & $-0.326 \pm 0.074$ \\
$24$ & $384$ & $26$ & $0.759$ & $0.411$ & $2.631 \pm 0.025$ & $0.368 \pm 0.057$  & $0.25507 \pm 0.00148$ & $0.2445 \pm 0.0258$ & $-0.2072 \pm 0.0784$ & $-$ & $0.555 \pm 0.030$ & $-0.267 \pm 0.092$ \\
$32$ & $384$ & $24$ & $0.755$ & $0.381$ & $2.611 \pm 0.025$ & $0.428 \pm 0.076$  & $0.25623 \pm 0.00143$ & $0.2863 \pm 0.0525$ & $-0.3417 \pm 0.1904$ & $-$ & $0.529 \pm 0.028$ & $-0.213 \pm 0.113$ \\
$36$ & $384$ & $22$ & $0.724$ & $0.387$ & $2.591 \pm 0.027$ & $0.450 \pm 0.101$  & $0.25656 \pm 0.00164$ & $0.3050 \pm 0.0824$ & $-0.4113 \pm 0.3231$ & $-$ & $0.508 \pm 0.028$ & $-0.123 \pm 0.133$ \\
\hline
$16$ & $256$ & $28$ & $0.810$ & $0.362$ & $2.667 \pm 0.019$ & $0.425 \pm 0.029$  & $0.25640 \pm 0.00068$ & $0.2733 \pm 0.0129$ & $-0.3029 \pm 0.0479$ & $-$ & $0.589 \pm 0.021$ & $-0.430 \pm 0.069$ \\
$20$ & $256$ & $26$ & $0.783$ & $0.374$ & $2.650 \pm 0.023$ & $0.427 \pm 0.045$  & $0.25644 \pm 0.00098$ & $0.2746 \pm 0.0228$ & $-0.3072 \pm 0.0825$ & $-$ & $0.571 \pm 0.025$ & $-0.369 \pm 0.085$ \\
$24$ & $256$ & $24$ & $0.770$ & $0.359$ & $2.640 \pm 0.029$ & $0.394 \pm 0.066$  & $0.25576 \pm 0.00155$ & $0.2558 \pm 0.0322$ & $-0.2442 \pm 0.1046$ & $-$ & $0.563 \pm 0.035$ & $-0.310 \pm 0.109$ \\
$32$ & $256$ & $22$ & $0.744$ & $0.358$ & $2.615 \pm 0.029$ & $0.496 \pm 0.085$  & $0.25743 \pm 0.00126$ & $0.3331 \pm 0.0710$ & $-0.5371 \pm 0.3224$ & $-$ & $0.531 \pm 0.030$ & $-0.266 \pm 0.145$ \\
$36$ & $256$ & $20$ & $0.693$ & $0.384$ & $2.590 \pm 0.029$ & $0.569 \pm 0.106$  & $0.25826 \pm 0.00118$ & $0.4147 \pm 0.1256$ & $-0.9664 \pm 0.7638$ & $-$ & $0.504 \pm 0.027$ & $-0.154 \pm 0.186$ \\
\hline
$16$ & $192$ & $26$ & $0.859$ & $0.268$ & $2.670 \pm 0.020$ & $0.422 \pm 0.031$  & $0.25631 \pm 0.00073$ & $0.2721 \pm 0.0131$ & $-0.2976 \pm 0.0487$ & $-$ & $0.593 \pm 0.023$ & $-0.436 \pm 0.072$ \\
$20$ & $192$ & $24$ & $0.838$ & $0.268$ & $2.653 \pm 0.024$ & $0.421 \pm 0.049$  & $0.25628 \pm 0.00109$ & $0.2716 \pm 0.0235$ & $-0.2959 \pm 0.0844$ & $-$ & $0.574 \pm 0.028$ & $-0.374 \pm 0.089$ \\
$24$ & $192$ & $22$ & $0.824$ & $0.256$ & $2.644 \pm 0.033$ & $0.380 \pm 0.074$  & $0.25538 \pm 0.00187$ & $0.2496 \pm 0.0333$ & $-0.2235 \pm 0.1059$ & $-$ & $0.568 \pm 0.040$ & $-0.314 \pm 0.116$ \\
$32$ & $192$ & $20$ & $0.814$ & $0.234$ & $2.616 \pm 0.032$ & $0.485 \pm 0.094$  & $0.25726 \pm 0.00147$ & $0.3254 \pm 0.0748$ & $-0.5019 \pm 0.3295$ & $-$ & $0.532 \pm 0.033$ & $-0.262 \pm 0.153$ \\
$36$ & $192$ & $18$ & $0.766$ & $0.245$ & $2.587 \pm 0.032$ & $0.564 \pm 0.117$  & $0.25820 \pm 0.00134$ & $0.4090 \pm 0.1343$ & $-0.9314 \pm 0.8035$ & $-$ & $0.502 \pm 0.030$ & $-0.137 \pm 0.197$ \\
\hline\hline
$16$ & $384$ & $30$ & $0.874$ & $0.242$ & $2.651 \pm 0.036$ & $0.429 \pm 0.154$  & $0.25629 \pm 0.00210$ & $0.2871 \pm 0.1659$ & $-0.3668 \pm 0.7741$ & $0.0878 \pm 0.8625$ & $0.573 \pm 0.050$ & $-0.388 \pm 0.068$ \\
$16$ & $256$ & $28$ & $0.801$ & $0.318$ & $2.653 \pm 0.031$ & $0.490 \pm 0.133$  & $0.25722 \pm 0.00171$ & $0.3437 \pm 0.1457$ & $-0.6670 \pm 0.8674$ & $0.4580 \pm 1.2776$ & $0.570 \pm 0.038$ & $-0.436 \pm 0.070$ \\
$16$ & $192$ & $26$ & $0.854$ & $0.223$ & $2.657 \pm 0.038$ & $0.475 \pm 0.159$  & $0.25701 \pm 0.00212$ & $0.3291 \pm 0.1695$ & $-0.5825 \pm 0.9517$ & $0.3410 \pm 1.3033$ & $0.575 \pm 0.049$ & $-0.439 \pm 0.074$ \\
\hline\hline
\end{tabular}
\end{minipage}
}
\end{table*}



\begin{table*}
\rotatebox{90}{
\begin{minipage}{1\textheight}
\caption{
List of fitting parameters corresponding to the global minimum $\chi^2_\text{min}$ of the cost function, obtained by method I for $0\le x \le0.07$. $M_\text{min}$ and  $M_\text{max}$ represent the minimum and maximum circumferences, respectively, used in the analysis. $N$ and $Q$ indicate the total number of input data points and goodness of fit, respectively.
\label{t2}
}
\begin{tabular}{c c c|c c | c c c | c c c| c c}
\hline\hline
$M_\text{min}$ &  $M_\text{max}$ & $N$ & $\chi^2_\text{min}/N$ & $Q$
 &$\nu$ & $|y|$ & $\Gamma_c/\pi \equiv \alpha_0-d$ &
$a_1$ &$a_2$  &$a_3$ & $\Gamma^\prime$ & $a_1^\prime$\\
\hline
$24$ & $384$ & $26$ & $1.075$ & $0.111$ & $2.669 \pm 0.027$ & $0.137 \pm 0.012$  & $0.24354 \pm 0.00202$ & $0.1637 \pm 0.0078$ & $-$ & $-$ & $0.692 \pm 0.066$ & $-0.369 \pm 0.105$ \\
$32$ & $384$ & $24$ & $0.868$ & $0.288$ & $2.628 \pm 0.028$ & $0.163 \pm 0.020$  & $0.24702 \pm 0.00230$ & $0.1518 \pm 0.0072$ & $-$ & $-$ & $0.597 \pm 0.056$ & $-0.249 \pm 0.107$ \\
$36$ & $384$ & $22$ & $0.855$ & $0.278$ & $2.618 \pm 0.029$ & $0.182 \pm 0.024$  & $0.24891 \pm 0.00223$ & $0.1468 \pm 0.0057$ & $-$ & $-$ & $0.574 \pm 0.052$ & $-0.218 \pm 0.108$ \\
$40$ & $384$ & $20$ & $0.817$ & $0.293$ & $2.604 \pm 0.030$ & $0.200 \pm 0.028$  & $0.25039 \pm 0.00209$ & $0.1441 \pm 0.0039$ & $-$ & $-$ & $0.547 \pm 0.049$ & $-0.170 \pm 0.113$ \\
$48$ & $384$ & $18$ & $0.647$ & $0.474$ & $2.566 \pm 0.031$ & $0.243 \pm 0.043$  & $0.25281 \pm 0.00210$ & $0.1445 \pm 0.0025$ & $-$ & $-$ & $0.489 \pm 0.041$ & $-0.034 \pm 0.121$ \\
$64$ & $384$ & $16$ & $0.722$ & $0.316$ & $2.570 \pm 0.040$ & $0.255 \pm 0.064$  & $0.25333 \pm 0.00275$ & $0.1457 \pm 0.0067$ & $-$ & $-$ & $0.495 \pm 0.053$ & $-0.051 \pm 0.161$ \\
\hline
$24$ & $256$ & $24$ & $0.817$ & $0.355$ & $2.716 \pm 0.034$ & $0.132 \pm 0.013$  & $0.24271 \pm 0.00234$ & $0.1670 \pm 0.0093$ & $-$ & $-$ & $0.793 \pm 0.098$ & $-0.469 \pm 0.133$ \\
$32$ & $256$ & $22$ & $0.729$ & $0.450$ & $2.675 \pm 0.039$ & $0.158 \pm 0.023$  & $0.24647 \pm 0.00278$ & $0.1537 \pm 0.0091$ & $-$ & $-$ & $0.682 \pm 0.089$ & $-0.368 \pm 0.145$ \\
$36$ & $256$ & $20$ & $0.707$ & $0.439$ & $2.670 \pm 0.041$ & $0.181 \pm 0.029$  & $0.24882 \pm 0.00265$ & $0.1471 \pm 0.0069$ & $-$ & $-$ & $0.657 \pm 0.084$ & $-0.352 \pm 0.147$ \\
$40$ & $256$ & $18$ & $0.674$ & $0.435$ & $2.659 \pm 0.044$ & $0.205 \pm 0.034$  & $0.25076 \pm 0.00240$ & $0.1434 \pm 0.0041$ & $-$ & $-$ & $0.627 \pm 0.079$ & $-0.323 \pm 0.154$ \\
$48$ & $256$ & $16$ & $0.558$ & $0.539$ & $2.612 \pm 0.046$ & $0.272 \pm 0.054$  & $0.25410 \pm 0.00215$ & $0.1462 \pm 0.0060$ & $-$ & $-$ & $0.546 \pm 0.062$ & $-0.194 \pm 0.168$ \\
$64$ & $256$ & $14$ & $0.446$ & $0.620$ & $2.672 \pm 0.070$ & $0.334 \pm 0.093$  & $0.25599 \pm 0.00233$ & $0.1591 \pm 0.0239$ & $-$ & $-$ & $0.616 \pm 0.100$ & $-0.433 \pm 0.258$ \\
\hline
$24$ & $192$ & $22$ & $0.788$ & $0.364$ & $2.725 \pm 0.037$ & $0.127 \pm 0.013$  & $0.24185 \pm 0.00258$ & $0.1705 \pm 0.0105$ & $-$ & $-$ & $0.819 \pm 0.112$ & $-0.488 \pm 0.146$ \\
$32$ & $192$ & $20$ & $0.731$ & $0.405$ & $2.684 \pm 0.044$ & $0.150 \pm 0.024$  & $0.24543 \pm 0.00323$ & $0.1573 \pm 0.0112$ & $-$ & $-$ & $0.705 \pm 0.106$ & $-0.391 \pm 0.166$ \\
$36$ & $192$ & $18$ & $0.728$ & $0.362$ & $2.680 \pm 0.047$ & $0.171 \pm 0.030$  & $0.24789 \pm 0.00313$ & $0.1497 \pm 0.0090$ & $-$ & $-$ & $0.681 \pm 0.102$ & $-0.377 \pm 0.169$ \\
$40$ & $192$ & $16$ & $0.713$ & $0.327$ & $2.669 \pm 0.050$ & $0.196 \pm 0.036$  & $0.25004 \pm 0.00282$ & $0.1449 \pm 0.0058$ & $-$ & $-$ & $0.648 \pm 0.096$ & $-0.349 \pm 0.176$ \\
$48$ & $192$ & $14$ & $0.624$ & $0.366$ & $2.617 \pm 0.053$ & $0.262 \pm 0.059$  & $0.25370 \pm 0.00250$ & $0.1454 \pm 0.0051$ & $-$ & $-$ & $0.554 \pm 0.075$ & $-0.209 \pm 0.193$ \\
$64$ & $192$ & $12$ & $0.447$ & $0.498$ & $2.717 \pm 0.094$ & $0.325 \pm 0.104$  & $0.25574 \pm 0.00279$ & $0.1569 \pm 0.0247$ & $-$ & $-$ & $0.676 \pm 0.149$ & $-0.558 \pm 0.322$ \\
\hline\hline
$16$ & $384$ & $30$ & $1.352$ & $0.013$ & $2.623 \pm 0.008$ & $0.419 \pm 0.027$  & $0.25624 \pm 0.00063$ & $0.2716 \pm 0.0117$ & $-0.2951 \pm 0.0427$ & $-$ & $0.549 \pm 0.010$ & $-0.311 \pm 0.029$ \\
$20$ & $384$ & $28$ & $0.935$ & $0.200$ & $2.609 \pm 0.010$ & $0.410 \pm 0.040$  & $0.25602 \pm 0.00092$ & $0.2672 \pm 0.0191$ & $-0.2790 \pm 0.0663$ & $-$ & $0.534 \pm 0.012$ & $-0.250 \pm 0.035$ \\
$24$ & $384$ & $26$ & $0.790$ & $0.363$ & $2.603 \pm 0.013$ & $0.374 \pm 0.056$  & $0.25523 \pm 0.00142$ & $0.2473 \pm 0.0260$ & $-0.2158 \pm 0.0799$ & $-$ & $0.529 \pm 0.016$ & $-0.206 \pm 0.044$ \\
$32$ & $384$ & $24$ & $0.674$ & $0.512$ & $2.585 \pm 0.012$ & $0.432 \pm 0.075$  & $0.25629 \pm 0.00140$ & $0.2886 \pm 0.0526$ & $-0.3501 \pm 0.1930$ & $-$ & $0.508 \pm 0.014$ & $-0.151 \pm 0.053$ \\
$36$ & $384$ & $22$ & $0.728$ & $0.381$ & $2.583 \pm 0.013$ & $0.459 \pm 0.098$  & $0.25671 \pm 0.00154$ & $0.3129 \pm 0.0830$ & $-0.4426 \pm 0.3374$ & $-$ & $0.505 \pm 0.014$ & $-0.149 \pm 0.062$ \\
\hline
$16$ & $256$ & $28$ & $0.909$ & $0.228$ & $2.638 \pm 0.010$ & $0.429 \pm 0.029$  & $0.25648 \pm 0.00067$ & $0.2748 \pm 0.0130$ & $-0.3085 \pm 0.0484$ & $-$ & $0.562 \pm 0.012$ & $-0.360 \pm 0.033$ \\
$20$ & $256$ & $26$ & $0.628$ & $0.635$ & $2.623 \pm 0.012$ & $0.429 \pm 0.045$  & $0.25647 \pm 0.00097$ & $0.2753 \pm 0.0228$ & $-0.3098 \pm 0.0829$ & $-$ & $0.546 \pm 0.014$ & $-0.302 \pm 0.040$ \\
$24$ & $256$ & $24$ & $0.569$ & $0.692$ & $2.619 \pm 0.016$ & $0.394 \pm 0.066$  & $0.25576 \pm 0.00155$ & $0.2558 \pm 0.0321$ & $-0.2443 \pm 0.1044$ & $-$ & $0.544 \pm 0.021$ & $-0.267 \pm 0.052$ \\
$32$ & $256$ & $22$ & $0.493$ & $0.763$ & $2.598 \pm 0.015$ & $0.492 \pm 0.086$  & $0.25737 \pm 0.00129$ & $0.3301 \pm 0.0707$ & $-0.5237 \pm 0.3169$ & $-$ & $0.518 \pm 0.016$ & $-0.227 \pm 0.067$ \\
$36$ & $256$ & $20$ & $0.496$ & $0.701$ & $2.595 \pm 0.015$ & $0.568 \pm 0.106$  & $0.25825 \pm 0.00119$ & $0.4128 \pm 0.1252$ & $-0.9547 \pm 0.7565$ & $-$ & $0.514 \pm 0.015$ & $-0.260 \pm 0.087$ \\
\hline
$16$ & $192$ & $26$ & $0.939$ & $0.181$ & $2.641 \pm 0.011$ & $0.425 \pm 0.030$  & $0.25638 \pm 0.00071$ & $0.2734 \pm 0.0132$ & $-0.3027 \pm 0.0492$ & $-$ & $0.566 \pm 0.013$ & $-0.367 \pm 0.034$ \\
$20$ & $192$ & $24$ & $0.658$ & $0.539$ & $2.626 \pm 0.013$ & $0.422 \pm 0.049$  & $0.25630 \pm 0.00108$ & $0.2721 \pm 0.0235$ & $-0.2975 \pm 0.0846$ & $-$ & $0.550 \pm 0.016$ & $-0.308 \pm 0.042$ \\
$24$ & $192$ & $22$ & $0.595$ & $0.595$ & $2.624 \pm 0.019$ & $0.379 \pm 0.074$  & $0.25536 \pm 0.00188$ & $0.2492 \pm 0.0333$ & $-0.2222 \pm 0.1054$ & $-$ & $0.550 \pm 0.025$ & $-0.275 \pm 0.057$ \\
$32$ & $192$ & $20$ & $0.535$ & $0.636$ & $2.600 \pm 0.017$ & $0.480 \pm 0.095$  & $0.25718 \pm 0.00152$ & $0.3214 \pm 0.0743$ & $-0.4843 \pm 0.3218$ & $-$ & $0.521 \pm 0.018$ & $-0.232 \pm 0.070$ \\
$36$ & $192$ & $18$ & $0.546$ & $0.546$ & $2.597 \pm 0.017$ & $0.561 \pm 0.118$  & $0.25816 \pm 0.00136$ & $0.4051 \pm 0.1336$ & $-0.9086 \pm 0.7892$ & $-$ & $0.516 \pm 0.017$ & $-0.266 \pm 0.092$ \\
\hline\hline
$16$ & $384$ & $30$ & $1.318$ & $0.012$ & $2.612 \pm 0.013$ & $0.499 \pm 0.090$  & $0.25722 \pm 0.00115$ & $0.3614 \pm 0.1024$ & $-0.7728 \pm 0.6469$ & $0.6350 \pm 1.0337$ & $0.533 \pm 0.016$ & $-0.316 \pm 0.029$ \\
$16$ & $256$ & $28$ & $0.877$ & $0.219$ & $2.621 \pm 0.016$ & $0.527 \pm 0.103$  & $0.25769 \pm 0.00126$ & $0.3854 \pm 0.1199$ & $-0.9358 \pm 0.8326$ & $0.8907 \pm 1.4676$ & $0.540 \pm 0.019$ & $-0.368 \pm 0.033$ \\
$16$ & $192$ & $26$ & $0.914$ & $0.163$ & $2.624 \pm 0.020$ & $0.516 \pm 0.118$  & $0.25754 \pm 0.00149$ & $0.3730 \pm 0.1330$ & $-0.8515 \pm 0.8845$ & $0.7481 \pm 1.4785$ & $0.543 \pm 0.024$ & $-0.372 \pm 0.034$ \\
\hline\hline
\end{tabular}
\end{minipage}
}
\end{table*}



\begin{table*}
\rotatebox{90}{
\begin{minipage}{1\textheight}
\caption{
List of fitting parameters corresponding to the global minimum $\chi^2_\text{min}$ of the cost function, obtained by method II for $0\le x \le0.05$. $M_\text{min}$ and  $M_\text{max}$ represent the minimum and maximum circumferences, respectively, used in the analysis. $N$ and $Q$ indicate the total number of input data points and goodness of fit, respectively.
\label{t3}
}
\begin{tabular}{c c c|c c | c c c | c c c| c c}
\hline\hline
$M_\text{min}$ &  $M_\text{max}$ & $N$ & $\chi^2_\text{min}/N$ & $Q$
 &$\nu$ & $|y|$ & $\Gamma_c/\pi \equiv \alpha_0-d$ &
$a_1$ &$a_2$  &$a_3$ & $\Gamma^\prime$ & $a_1^\prime$\\
\hline
$32$ & $384$ & $97$ & $1.188$ & $0.044$ & $2.584 \pm 0.054$ & $0.157 \pm 0.007$  & $0.24629 \pm 0.00084$ & $0.1544 \pm 0.0029$ & $-$ & $-$ & $0.516 \pm 0.089$ & $-0.084 \pm 0.192$ \\
$36$ & $384$ & $86$ & $1.067$ & $0.173$ & $2.575 \pm 0.058$ & $0.173 \pm 0.008$  & $0.24806 \pm 0.00082$ & $0.1491 \pm 0.0024$ & $-$ & $-$ & $0.498 \pm 0.089$ & $-0.039 \pm 0.214$ \\
$40$ & $384$ & $75$ & $0.631$ & $0.978$ & $2.586 \pm 0.056$ & $0.202 \pm 0.009$  & $0.25057 \pm 0.00070$ & $0.1437 \pm 0.0014$ & $-$ & $-$ & $0.509 \pm 0.080$ & $-0.063 \pm 0.199$ \\
$48$ & $384$ & $64$ & $0.680$ & $0.921$ & $2.601 \pm 0.062$ & $0.217 \pm 0.013$  & $0.25150 \pm 0.00084$ & $0.1428 \pm 0.0010$ & $-$ & $-$ & $0.528 \pm 0.088$ & $-0.110 \pm 0.217$ \\
$64$ & $384$ & $53$ & $0.660$ & $0.902$ & $2.729 \pm 0.060$ & $0.260 \pm 0.030$  & $0.25360 \pm 0.00126$ & $0.1451 \pm 0.0029$ & $-$ & $-$ & $0.713 \pm 0.088$ & $-0.529 \pm 0.168$ \\
\hline
$32$ & $256$ & $91$ & $1.236$ & $0.025$ & $2.503 \pm 0.110$ & $0.152 \pm 0.008$  & $0.24572 \pm 0.00103$ & $0.1564 \pm 0.0036$ & $-$ & $-$ & $0.391 \pm 0.164$ & $0.258 \pm 0.580$ \\
$36$ & $256$ & $80$ & $1.102$ & $0.125$ & $2.447 \pm 0.156$ & $0.168 \pm 0.009$  & $0.24756 \pm 0.00100$ & $0.1505 \pm 0.0031$ & $-$ & $-$ & $0.317 \pm 0.204$ & $0.598 \pm 1.092$ \\
$40$ & $256$ & $69$ & $0.595$ & $0.986$ & $2.447 \pm 0.143$ & $0.200 \pm 0.011$  & $0.25037 \pm 0.00085$ & $0.1439 \pm 0.0018$ & $-$ & $-$ & $0.329 \pm 0.171$ & $0.584 \pm 0.908$ \\
$48$ & $256$ & $58$ & $0.606$ & $0.965$ & $2.405 \pm 0.212$ & $0.215 \pm 0.016$  & $0.25143 \pm 0.00102$ & $0.1425 \pm 0.0013$ & $-$ & $-$ & $0.286 \pm 0.231$ & $0.878 \pm 1.630$ \\
$64$ & $256$ & $47$ & $0.522$ & $0.981$ & $2.743 \pm 0.105$ & $0.333 \pm 0.044$  & $0.25606 \pm 0.00112$ & $0.1565 \pm 0.0103$ & $-$ & $-$ & $0.705 \pm 0.138$ & $-0.631 \pm 0.334$ \\
\hline
$32$ & $192$ & $85$ & $1.241$ & $0.025$ & $2.494 \pm 0.130$ & $0.149 \pm 0.008$  & $0.24523 \pm 0.00113$ & $0.1582 \pm 0.0041$ & $-$ & $-$ & $0.377 \pm 0.193$ & $0.305 \pm 0.725$ \\
$40$ & $192$ & $63$ & $0.583$ & $0.983$ & $2.392 \pm 0.239$ & $0.196 \pm 0.012$  & $0.25005 \pm 0.00094$ & $0.1446 \pm 0.0021$ & $-$ & $-$ & $0.264 \pm 0.264$ & $0.993 \pm 2.075$ \\
$64$ & $192$ & $41$ & $0.514$ & $0.970$ & $2.760 \pm 0.131$ & $0.342 \pm 0.050$  & $0.25629 \pm 0.00121$ & $0.1586 \pm 0.0124$ & $-$ & $-$ & $0.726 \pm 0.171$ & $-0.690 \pm 0.409$ \\
\hline\hline
$16$ & $384$ & $130$ & $0.660$ & $0.996$ & $2.655 \pm 0.013$ & $0.428 \pm 0.008$  & $0.25641 \pm 0.00021$ & $0.2766 \pm 0.0033$ & $-0.3119 \pm 0.0131$ & $-$ & $0.578 \pm 0.014$ & $-0.400 \pm 0.048$ \\
$20$ & $384$ & $119$ & $0.689$ & $0.985$ & $2.644 \pm 0.015$ & $0.434 \pm 0.012$  & $0.25653 \pm 0.00027$ & $0.2800 \pm 0.0058$ & $-0.3244 \pm 0.0220$ & $-$ & $0.565 \pm 0.016$ & $-0.357 \pm 0.059$ \\
$24$ & $384$ & $108$ & $0.716$ & $0.962$ & $2.632 \pm 0.019$ & $0.412 \pm 0.017$  & $0.25608 \pm 0.00039$ & $0.2670 \pm 0.0086$ & $-0.2792 \pm 0.0297$ & $-$ & $0.553 \pm 0.020$ & $-0.294 \pm 0.074$ \\
$32$ & $384$ & $97$ & $0.728$ & $0.935$ & $2.629 \pm 0.020$ & $0.471 \pm 0.027$  & $0.25709 \pm 0.00044$ & $0.3102 \pm 0.0199$ & $-0.4386 \pm 0.0821$ & $-$ & $0.546 \pm 0.020$ & $-0.315 \pm 0.098$ \\
$36$ & $384$ & $86$ & $0.795$ & $0.797$ & $2.620 \pm 0.022$ & $0.482 \pm 0.035$  & $0.25724 \pm 0.00053$ & $0.3202 \pm 0.0291$ & $-0.4801 \pm 0.1244$ & $-$ & $0.536 \pm 0.023$ & $-0.274 \pm 0.119$ \\
\hline
$16$ & $256$ & $124$ & $0.621$ & $0.998$ & $2.655 \pm 0.016$ & $0.436 \pm 0.009$  & $0.25662 \pm 0.00022$ & $0.2790 \pm 0.0035$ & $-0.3231 \pm 0.0142$ & $-$ & $0.577 \pm 0.016$ & $-0.403 \pm 0.056$ \\
$20$ & $256$ & $113$ & $0.631$ & $0.996$ & $2.640 \pm 0.018$ & $0.448 \pm 0.013$  & $0.25684 \pm 0.00027$ & $0.2858 \pm 0.0063$ & $-0.3485 \pm 0.0248$ & $-$ & $0.560 \pm 0.019$ & $-0.350 \pm 0.072$ \\
$24$ & $256$ & $102$ & $0.667$ & $0.983$ & $2.623 \pm 0.023$ & $0.430 \pm 0.018$  & $0.25650 \pm 0.00039$ & $0.2757 \pm 0.0095$ & $-0.3115 \pm 0.0346$ & $-$ & $0.542 \pm 0.024$ & $-0.268 \pm 0.094$ \\
$32$ & $256$ & $91$ & $0.573$ & $0.998$ & $2.612 \pm 0.025$ & $0.526 \pm 0.027$  & $0.25793 \pm 0.00038$ & $0.3530 \pm 0.0241$ & $-0.6359 \pm 0.1197$ & $-$ & $0.527 \pm 0.024$ & $-0.270 \pm 0.135$ \\
$36$ & $256$ & $80$ & $0.551$ & $0.997$ & $2.596 \pm 0.027$ & $0.582 \pm 0.035$  & $0.25853 \pm 0.00038$ & $0.4158 \pm 0.0395$ & $-0.9843 \pm 0.2447$ & $-$ & $0.510 \pm 0.025$ & $-0.200 \pm 0.177$ \\
\hline
$16$ & $192$ & $118$ & $0.621$ & $0.998$ & $2.658 \pm 0.017$ & $0.436 \pm 0.009$  & $0.25661 \pm 0.00023$ & $0.2788 \pm 0.0036$ & $-0.3222 \pm 0.0145$ & $-$ & $0.579 \pm 0.017$ & $-0.410 \pm 0.058$ \\
$20$ & $192$ & $107$ & $0.634$ & $0.994$ & $2.642 \pm 0.020$ & $0.447 \pm 0.013$  & $0.25683 \pm 0.00029$ & $0.2855 \pm 0.0065$ & $-0.3475 \pm 0.0255$ & $-$ & $0.561 \pm 0.020$ & $-0.355 \pm 0.075$ \\
$24$ & $192$ & $96$ & $0.673$ & $0.976$ & $2.623 \pm 0.025$ & $0.429 \pm 0.019$  & $0.25647 \pm 0.00041$ & $0.2749 \pm 0.0097$ & $-0.3086 \pm 0.0354$ & $-$ & $0.543 \pm 0.026$ & $-0.269 \pm 0.099$ \\
$32$ & $192$ & $85$ & $0.572$ & $0.996$ & $2.610 \pm 0.027$ & $0.529 \pm 0.028$  & $0.25797 \pm 0.00039$ & $0.3557 \pm 0.0251$ & $-0.6494 \pm 0.1263$ & $-$ & $0.525 \pm 0.026$ & $-0.259 \pm 0.144$ \\
$36$ & $192$ & $74$ & $0.536$ & $0.997$ & $2.590 \pm 0.029$ & $0.592 \pm 0.036$  & $0.25864 \pm 0.00039$ & $0.4273 \pm 0.0421$ & $-1.0573 \pm 0.2710$ & $-$ & $0.505 \pm 0.027$ & $-0.169 \pm 0.193$ \\
\hline\hline
$16$ & $384$ & $130$ & $0.657$ & $0.995$ & $2.651 \pm 0.013$ & $0.461 \pm 0.045$  & $0.25683 \pm 0.00059$ & $0.3122 \pm 0.0487$ & $-0.4850 \pm 0.2548$ & $0.1989 \pm 0.3207$ & $0.570 \pm 0.015$ & $-0.410 \pm 0.052$ \\
$16$ & $256$ & $124$ & $0.583$ & $0.999$ & $2.644 \pm 0.014$ & $0.529 \pm 0.033$  & $0.25776 \pm 0.00040$ & $0.3829 \pm 0.0382$ & $-0.9177 \pm 0.2648$ & $0.8419 \pm 0.4628$ & $0.559 \pm 0.014$ & $-0.438 \pm 0.062$ \\
$16$ & $192$ & $118$ & $0.584$ & $0.999$ & $2.645 \pm 0.015$ & $0.530 \pm 0.034$  & $0.25777 \pm 0.00042$ & $0.3832 \pm 0.0395$ & $-0.9203 \pm 0.2749$ & $0.8461 \pm 0.4808$ & $0.560 \pm 0.014$ & $-0.443 \pm 0.064$ \\
\hline\hline
\end{tabular}
\end{minipage}
}
\end{table*}

\begin{table*}
\rotatebox{90}{
\begin{minipage}{1\textheight}
\caption{
List of fitting parameters corresponding to the global minimum $\chi^2_\text{min}$ of the cost function, obtained by method II for $0\le x \le0.07$. $M_\text{min}$ and  $M_\text{max}$ represent the minimum and maximum circumferences, respectively, used in the analysis. $N$ and $Q$ indicate the total number of input data points and goodness of fit, respectively.
\label{t4}
}
\begin{tabular}{c c c|c c | c c c | c c c| c c}
\hline\hline
$M_\text{min}$ &  $M_\text{max}$ & $N$ & $\chi^2_\text{min}/N$ & $Q$
 &$\nu$ & $|y|$ & $\Gamma_c/\pi \equiv \alpha_0-d$ &
$a_1$ &$a_2$  &$a_3$ & $\Gamma^\prime$ & $a_1^\prime$\\
\hline
$32$ & $384$ & $131$ & $1.225$ & $0.018$ & $2.588 \pm 0.025$ & $0.146 \pm 0.006$  & $0.24477 \pm 0.00085$ & $0.1601 \pm 0.0031$ & $-$ & $-$ & $0.535 \pm 0.043$ & $-0.134 \pm 0.086$ \\
$36$ & $384$ & $116$ & $1.214$ & $0.025$ & $2.595 \pm 0.026$ & $0.160 \pm 0.007$  & $0.24655 \pm 0.00083$ & $0.1543 \pm 0.0027$ & $-$ & $-$ & $0.542 \pm 0.043$ & $-0.150 \pm 0.085$ \\
$40$ & $384$ & $101$ & $0.904$ & $0.589$ & $2.606 \pm 0.025$ & $0.188 \pm 0.008$  & $0.24934 \pm 0.00071$ & $0.1471 \pm 0.0017$ & $-$ & $-$ & $0.553 \pm 0.038$ & $-0.180 \pm 0.080$ \\
$48$ & $384$ & $86$ & $0.936$ & $0.463$ & $2.593 \pm 0.030$ & $0.203 \pm 0.011$  & $0.25045 \pm 0.00083$ & $0.1451 \pm 0.0014$ & $-$ & $-$ & $0.529 \pm 0.044$ & $-0.128 \pm 0.103$ \\
$64$ & $384$ & $71$ & $1.066$ & $0.172$ & $2.617 \pm 0.036$ & $0.218 \pm 0.026$  & $0.25143 \pm 0.00154$ & $0.1446 \pm 0.0009$ & $-$ & $-$ & $0.563 \pm 0.052$ & $-0.212 \pm 0.121$ \\
\hline
$32$ & $256$ & $123$ & $1.227$ & $0.019$ & $2.600 \pm 0.034$ & $0.147 \pm 0.007$  & $0.24495 \pm 0.00096$ & $0.1595 \pm 0.0035$ & $-$ & $-$ & $0.555 \pm 0.059$ & $-0.171 \pm 0.108$ \\
$36$ & $256$ & $108$ & $1.211$ & $0.029$ & $2.611 \pm 0.036$ & $0.164 \pm 0.008$  & $0.24696 \pm 0.00093$ & $0.1529 \pm 0.0030$ & $-$ & $-$ & $0.567 \pm 0.058$ & $-0.197 \pm 0.108$ \\
$40$ & $256$ & $93$ & $0.844$ & $0.731$ & $2.627 \pm 0.036$ & $0.197 \pm 0.010$  & $0.25005 \pm 0.00077$ & $0.1454 \pm 0.0017$ & $-$ & $-$ & $0.581 \pm 0.053$ & $-0.237 \pm 0.103$ \\
$48$ & $256$ & $78$ & $0.840$ & $0.692$ & $2.602 \pm 0.048$ & $0.216 \pm 0.014$  & $0.25138 \pm 0.00090$ & $0.1436 \pm 0.0011$ & $-$ & $-$ & $0.539 \pm 0.067$ & $-0.154 \pm 0.156$ \\
$64$ & $256$ & $63$ & $0.866$ & $0.567$ & $2.691 \pm 0.051$ & $0.307 \pm 0.038$  & $0.25529 \pm 0.00115$ & $0.1520 \pm 0.0072$ & $-$ & $-$ & $0.646 \pm 0.065$ & $-0.469 \pm 0.171$ \\
\hline
$32$ & $192$ & $115$ & $1.217$ & $0.024$ & $2.606 \pm 0.037$ & $0.145 \pm 0.007$  & $0.24461 \pm 0.00104$ & $0.1608 \pm 0.0039$ & $-$ & $-$ & $0.567 \pm 0.064$ & $-0.191 \pm 0.113$ \\
$36$ & $192$ & $100$ & $1.211$ & $0.031$ & $2.618 \pm 0.039$ & $0.162 \pm 0.008$  & $0.24673 \pm 0.00101$ & $0.1537 \pm 0.0033$ & $-$ & $-$ & $0.580 \pm 0.064$ & $-0.220 \pm 0.113$ \\
$40$ & $192$ & $85$ & $0.827$ & $0.746$ & $2.636 \pm 0.039$ & $0.196 \pm 0.010$  & $0.24999 \pm 0.00083$ & $0.1456 \pm 0.0018$ & $-$ & $-$ & $0.595 \pm 0.058$ & $-0.262 \pm 0.109$ \\
$48$ & $192$ & $70$ & $0.830$ & $0.684$ & $2.607 \pm 0.054$ & $0.215 \pm 0.015$  & $0.25133 \pm 0.00098$ & $0.1437 \pm 0.0012$ & $-$ & $-$ & $0.547 \pm 0.077$ & $-0.172 \pm 0.173$ \\
$64$ & $192$ & $55$ & $0.841$ & $0.585$ & $2.731 \pm 0.060$ & $0.327 \pm 0.043$  & $0.25585 \pm 0.00116$ & $0.1559 \pm 0.0097$ & $-$ & $-$ & $0.694 \pm 0.076$ & $-0.601 \pm 0.200$ \\
\hline\hline
$16$ & $384$ & $176$ & $0.820$ & $0.915$ & $2.624 \pm 0.006$ & $0.425 \pm 0.007$  & $0.25623 \pm 0.00019$ & $0.2791 \pm 0.0028$ & $-0.3149 \pm 0.0111$ & $-$ & $0.549 \pm 0.006$ & $-0.317 \pm 0.023$ \\
$20$ & $384$ & $161$ & $0.810$ & $0.916$ & $2.612 \pm 0.007$ & $0.425 \pm 0.010$  & $0.25623 \pm 0.00025$ & $0.2798 \pm 0.0048$ & $-0.3173 \pm 0.0183$ & $-$ & $0.536 \pm 0.007$ & $-0.266 \pm 0.029$ \\
$24$ & $384$ & $146$ & $0.830$ & $0.860$ & $2.604 \pm 0.009$ & $0.400 \pm 0.015$  & $0.25567 \pm 0.00037$ & $0.2655 \pm 0.0071$ & $-0.2678 \pm 0.0243$ & $-$ & $0.528 \pm 0.009$ & $-0.219 \pm 0.036$ \\
$32$ & $384$ & $131$ & $0.873$ & $0.721$ & $2.595 \pm 0.010$ & $0.438 \pm 0.024$  & $0.25640 \pm 0.00046$ & $0.2921 \pm 0.0163$ & $-0.3602 \pm 0.0611$ & $-$ & $0.518 \pm 0.010$ & $-0.195 \pm 0.047$ \\
$36$ & $384$ & $116$ & $0.957$ & $0.429$ & $2.597 \pm 0.010$ & $0.464 \pm 0.031$  & $0.25682 \pm 0.00051$ & $0.3127 \pm 0.0240$ & $-0.4402 \pm 0.0983$ & $-$ & $0.519 \pm 0.011$ & $-0.216 \pm 0.056$ \\
\hline
$16$ & $256$ & $168$ & $0.706$ & $0.995$ & $2.633 \pm 0.007$ & $0.436 \pm 0.008$  & $0.25655 \pm 0.00019$ & $0.2819 \pm 0.0030$ & $-0.3297 \pm 0.0122$ & $-$ & $0.557 \pm 0.007$ & $-0.349 \pm 0.026$ \\
$20$ & $256$ & $153$ & $0.691$ & $0.995$ & $2.619 \pm 0.008$ & $0.443 \pm 0.011$  & $0.25666 \pm 0.00025$ & $0.2862 \pm 0.0053$ & $-0.3454 \pm 0.0209$ & $-$ & $0.541 \pm 0.008$ & $-0.296 \pm 0.034$ \\
$24$ & $256$ & $138$ & $0.727$ & $0.979$ & $2.610 \pm 0.011$ & $0.423 \pm 0.016$  & $0.25626 \pm 0.00035$ & $0.2749 \pm 0.0080$ & $-0.3040 \pm 0.0288$ & $-$ & $0.534 \pm 0.011$ & $-0.250 \pm 0.043$ \\
$32$ & $256$ & $123$ & $0.702$ & $0.982$ & $2.598 \pm 0.012$ & $0.497 \pm 0.025$  & $0.25745 \pm 0.00038$ & $0.3316 \pm 0.0199$ & $-0.5285 \pm 0.0900$ & $-$ & $0.518 \pm 0.011$ & $-0.229 \pm 0.062$ \\
$36$ & $256$ & $108$ & $0.704$ & $0.970$ & $2.599 \pm 0.012$ & $0.566 \pm 0.030$  & $0.25829 \pm 0.00036$ & $0.4021 \pm 0.0330$ & $-0.8916 \pm 0.1936$ & $-$ & $0.518 \pm 0.012$ & $-0.282 \pm 0.080$ \\
\hline
$16$ & $192$ & $160$ & $0.690$ & $0.996$ & $2.635 \pm 0.007$ & $0.437 \pm 0.008$  & $0.25656 \pm 0.00020$ & $0.2819 \pm 0.0030$ & $-0.3299 \pm 0.0125$ & $-$ & $0.559 \pm 0.007$ & $-0.356 \pm 0.027$ \\
$20$ & $192$ & $145$ & $0.677$ & $0.996$ & $2.621 \pm 0.009$ & $0.443 \pm 0.011$  & $0.25667 \pm 0.00025$ & $0.2863 \pm 0.0054$ & $-0.3460 \pm 0.0215$ & $-$ & $0.543 \pm 0.009$ & $-0.302 \pm 0.035$ \\
$24$ & $192$ & $130$ & $0.714$ & $0.980$ & $2.612 \pm 0.011$ & $0.422 \pm 0.016$  & $0.25625 \pm 0.00037$ & $0.2746 \pm 0.0082$ & $-0.3031 \pm 0.0297$ & $-$ & $0.535 \pm 0.011$ & $-0.255 \pm 0.045$ \\
$32$ & $192$ & $115$ & $0.686$ & $0.984$ & $2.598 \pm 0.013$ & $0.500 \pm 0.026$  & $0.25751 \pm 0.00040$ & $0.3344 \pm 0.0208$ & $-0.5415 \pm 0.0954$ & $-$ & $0.518 \pm 0.012$ & $-0.231 \pm 0.065$ \\
$36$ & $192$ & $100$ & $0.677$ & $0.978$ & $2.599 \pm 0.013$ & $0.577 \pm 0.031$  & $0.25843 \pm 0.00036$ & $0.4139 \pm 0.0352$ & $-0.9636 \pm 0.2158$ & $-$ & $0.517 \pm 0.012$ & $-0.287 \pm 0.086$ \\
\hline\hline
$16$ & $384$ & $176$ & $0.817$ & $0.911$ & $2.620 \pm 0.007$ & $0.456 \pm 0.039$  & $0.25663 \pm 0.00053$ & $0.3122 \pm 0.0419$ & $-0.4751 \pm 0.2170$ & $0.1817 \pm 0.2679$ & $0.543 \pm 0.009$ & $-0.322 \pm 0.025$ \\
$16$ & $256$ & $168$ & $0.680$ & $0.998$ & $2.622 \pm 0.007$ & $0.516 \pm 0.030$  & $0.25753 \pm 0.00038$ & $0.3704 \pm 0.0346$ & $-0.8216 \pm 0.2280$ & $0.6671 \pm 0.3728$ & $0.541 \pm 0.007$ & $-0.364 \pm 0.030$ \\
$16$ & $192$ & $160$ & $0.664$ & $0.998$ & $2.624 \pm 0.007$ & $0.517 \pm 0.032$  & $0.25756 \pm 0.00040$ & $0.3710 \pm 0.0359$ & $-0.8267 \pm 0.2372$ & $0.6746 \pm 0.3891$ & $0.543 \pm 0.008$ & $-0.371 \pm 0.031$ \\
\hline\hline
\end{tabular}
\end{minipage}
}
\end{table*}

\begin{table*}[h]
\caption{
In this table we compare our results for the exponents $\nu$, $|y|$, $\Gamma_c$, and $\alpha_0-2$ with those of other authors. In the second column the abbreviations ``CC model'', ``LLL'', and ``TBM'' stand for the Chalker-Coddington network model, the lowest Landau level basis approximation, and the tight-binding model in a perpendicular magnetic field, respectively. Also, systems with different dimensionality are denotes as ``2D'' (two dimensions), ``Q1D'' (quasi-one dimension), and ``1D'' (one dimension). ``PBC`` and ``RBC'' stand for the periodic and reflecting boundary conditions, respectively, in the transverse direction in Q1D.
} \label{tab:compare}
 \begin{tabular}{l| l|l l l l}
\hline\hline
  & model & $\nu$ & $|y|$ & $\Gamma_c/\pi$ & $\alpha_0-2$ \\
\hline
present work & CC model with PBC in Q1D
& $2.62\pm0.06$ &$\gapproxeq0.4$ & $0.257 \pm 0.002$ &  - \\
\hline
Huckestein {\it et al.} \cite{Huckestein90} & LLL in Q1D&
$2.34 \pm 0.04$ & - & - & -\\
Mieck \cite{Mieck90} & LLL in Q1D&
$2.3 \pm 0.08$ & - & - & -\\
Huckestein \cite{Huckestein92} & LLL in Q1D&
$2.33 \pm 0.05$ & - & - & -\\
Huo {\it et al.}\cite{Huo92} & TBM in 2D &
$2.4 \pm 0.1$ & - & - & -\\
Huckestein \cite{Huckestein94} & LLL in Q1D&
- & $0.38\pm0.04$ & $0.279\pm0.05$ & -\\
Lee {\it et al.}\cite{Lee96} & CC model in Q1D&
$2.33 \pm 0.03$ & - & - & -\\
Cain {\it et al.}\cite{Cain03} & CC model in 2D&
$2.37 \pm 0.02$ & - & - & -\\
Sandler {\it et al.} \cite{Sandler04}& LLL in 2D&
$2.33 \pm 0.09$ & - & - & - \\
Slevin {\it et al.} \cite{Slevin09} & CC model with PBC in Q1D&
$2.593 \pm 0.006\quad$ & $0.17$ & $0.248\pm0.004$ & -\\
Obuse {\it et al.} \cite{s:Obuse10} & CC model with RBC in Q1D&
$2.55\pm 0.01$ & $1.29$ & - & -\\
Amado {\it et al.} \cite{s:Amado11} & CC model with PBC in Q1D &
$2.616 \pm 0.014$ & $0$ (logarithmic correction)$\quad$ &
	      $0.223\pm0.05$ & -\\
Dahlhaus {\it et al.} \cite{s:Dahlhaus11}& kicked rotator in 1D &
$2.576\pm 0.03$ & - & - & -\\
Fulga {\it et al.} \cite{Fulga11} & CC model in Corbino disc&
$2.56 \pm 0.03$ & no corrections & - & -\\
Slevin {\it et al.} \cite{Slevin12} & CC model with PBC in Q1D&
$2.607 \pm 0.009$ & - & - & -\\
Janssen  \cite{Janssen98} & CC model in 2D &
- & - & - & $0.27 \pm 0.02$ \\
Evers {\it et al.} \cite{s:Evers01} & CC model in 2D &
- & - & - & $0.262 \pm 0.003$ \\
Obuse {\it et al.} \cite{s:Obuse08} & CC model in 2D &
- & - & - & $0.2617 \pm 0.0006$ \\
Evers {\it et al.} \cite{s:Evers08} & CC model in 2D &
- & - & - & $0.2596 \pm 0.0004$ \\
\hline\hline
 \end{tabular}
\end{table*}



\begin{thebibliography}{40}
\bibitem{fn2}
Irrelevant corrections to scaling play an important role in the work by L\"utken and Ross \cite{luetken06}. These authors postulate an anti-holomorphic symmetry property of the scaling function $\beta^z{=}\beta_{xy}{+}\ci\beta_{xx}$ in the coupling constant $\sigma{=}\sigma_{xy}{+}\ci\sigma_{xx}$, and use it to derive the relation $\nu = 10.42\ldots/c$ where $c$ is an unspecified integer \cite{luetken06, slevin12}. The choice $c=4$ would lead to values consistent with numerical estimates. A necessary condition for anti-holomorphic scaling is $|y| \nu = 1$. This condition is not inconsistent with our data in Fig. \ref{f3} (IIa,IIb), since the colored symbols correspond to values of $|y|\nu$ between 1.0 and 1.5 with pronounced clustering in the range 1.0-1.3. For this possibility, see also Ref.\ \onlinecite{slevin12}.
\bibitem{method-I}

Alternatively, we could have used the fitted data $\gamma_{M}$ instead of $\Gamma_{M}(0)$ for finding the optimal parameters of the functions $\gamma(M)$ and $\gamma'(M)$ (by modifying the cost function appropriately). Since both data sets, $\Gamma_{M}(0)$ and $\gamma_{M}$, hardly deviate from each other (see Fig. \ref{f2}), both procedures are equivalent in the present case.
\bibitem{fitting}
We have given trivial constraints to the fitting: $y$ is negative, contributions of correction terms become smaller than the main terms, $\Gamma_c$ and $\Gamma^\prime$, at the maximal $M$, and the goodness of fit is larger than 0.01. While we mainly discuss the case when $x$ from the interval $0 \le x \le 0.05$ ($\Gamma\apprle 1$) was used in the fitting, we have checked that our fits are stable against varying the range of $x$-values that we use \cite{suppl}.
\bibitem{error}
The estimation of error bars for fitting parameters is based on propagation of uncertainty. Therefore, these error bars represent 	volatility against the statistical fluctuation of raw data $\Gamma_M(x)$ within its error bar $\sigma_M(x)$.
\bibitem{a2=0}
Moreover, in method I, $a_2$ at the local minima of $\chi^2_\text{I}$ is almost zero, and the corresponding values of $|y|$ are located in $|y| \lapproxeq 0.2$, and $\Gamma_c/\pi \le 0.25$, as shown in Fig.\ \ref{f3} (I-b). This is consistent with the results of method II, when only the $a_1$ term is kept in $\gamma(M)$.
\bibitem{nu}
We estimate $\nu$ based on  data obtained by method II including second order correction terms in $\gamma(M)$. As can be seen from Fig. \ref{f3} (IIa) and also in Ref. \cite{suppl} (Tables III and IV) there is a small residual scattering of the fitting parameters $\nu$ with variation of the input data. Our estimate $\nu=2.62\pm 0.06$ exhibits error bars that describe the full width of the window of $\nu$-values that we have observed.
\end{thebibliography}
\end{document}